\input{aipcheck}

\documentclass[
    ,final            
  ]
  {aipproc}

\layoutstyle{6x9}

\def\De{\Delta}

\def\La{\Lambda}
\def\Si{\Sigma}

\def\to{\rightarrow}

\def\no{\nonumber}
\def\no{\nonumber}
\def\beq{\begin{eqnarray}}
\def\eeq{\end{eqnarray}}

\def\numu{\nu_\mu}
\def\numubar{\bar{\nu}_\mu}

\def\MBosc1POT{5.58\times 10^{20}}

\begin{document}
\title{MicroBooNE, A Liquid Argon Time Projection Chamber (LArTPC) Neutrino Experiment}

\classification{13.15.+g,14.60.Lm,95.55.Vj}
\keywords      {Liquid Argon, TPC, neutrino, 
MicroBooNE, cross section}

\author{Teppei Katori for the MicroBooNE collaboration}{
  address={Massachusetts Institute of Technology, Cambridge, MA}
}

\begin{abstract}
Liquid Argon time projection chamber (LArTPC) is 
a promising detector technology for future neutrino experiments. 
MicroBooNE is a upcoming LArTPC neutrino experiment 
which will be located on-axis of Booster Neutrino Beam (BNB) at Fermilab, USA.
The R\&D efforts on this detection method and 
related neutrino interaction measurements are discussed. 
\end{abstract}

\maketitle

\subsection{Liquid Argon Time Projection Chamber (LArTPC)}

\begin{figure}
\includegraphics[height=2.5in]{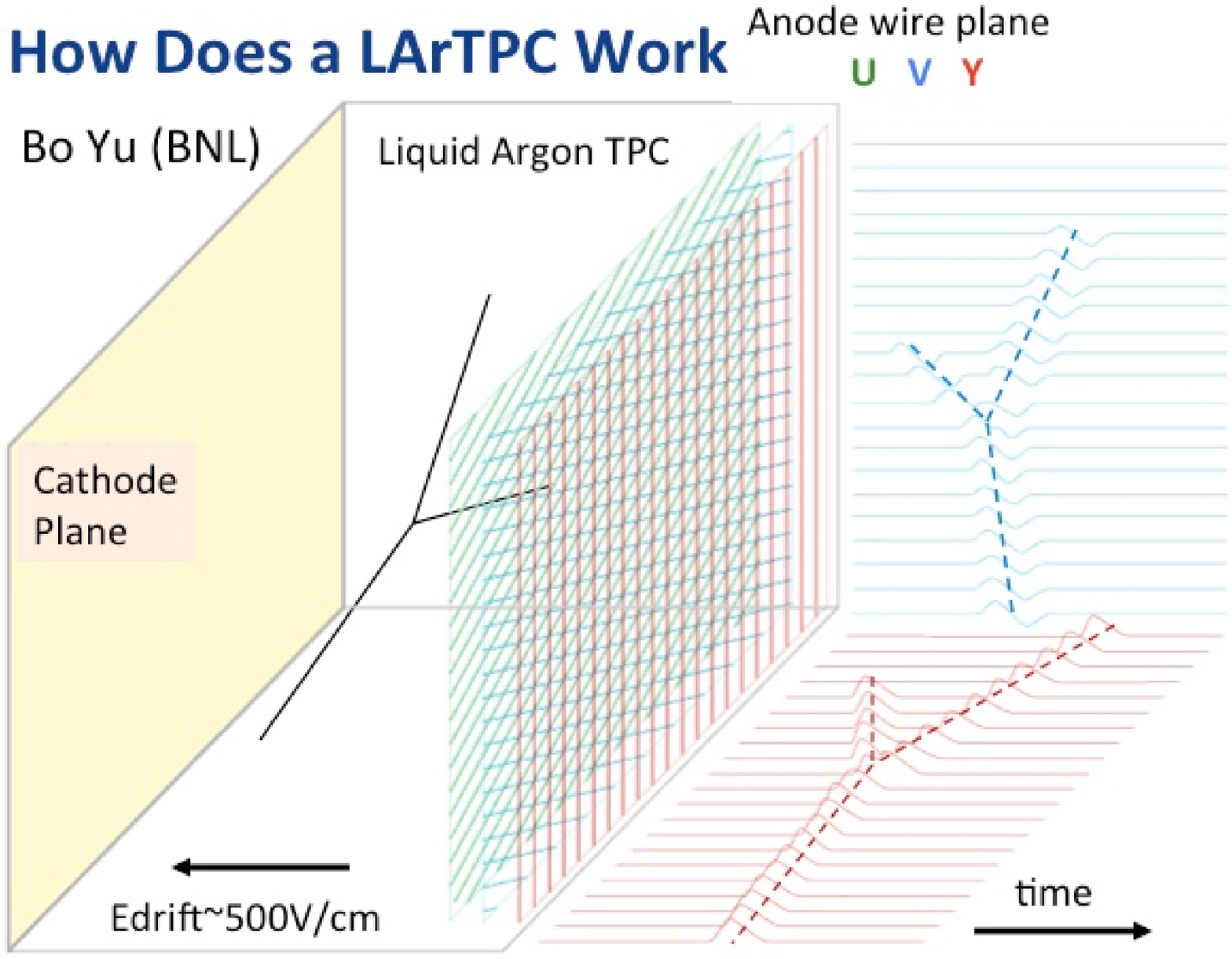}
\includegraphics[height=2.5in]{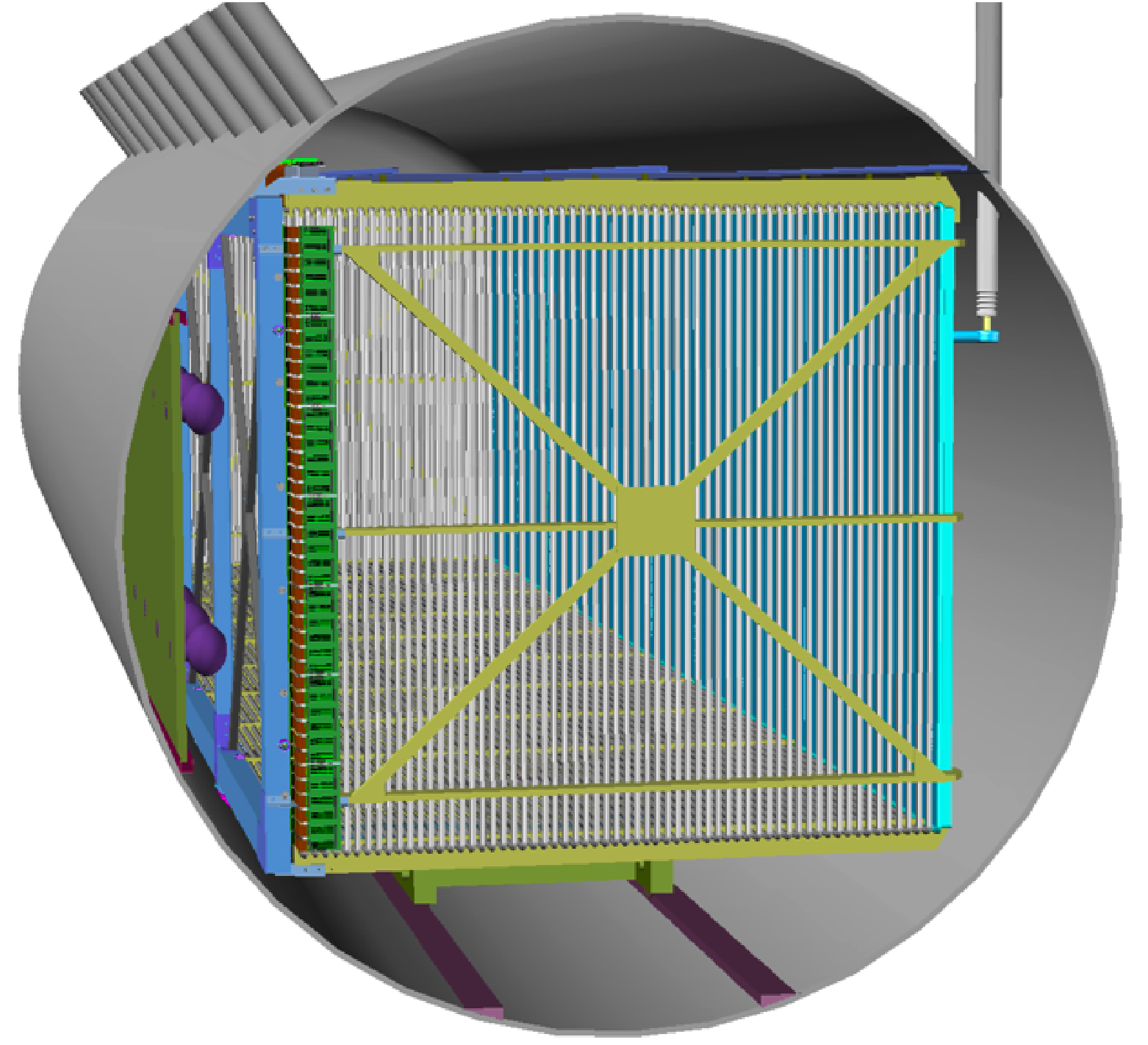}
\caption{
(color online). 
(Left) The working principle of a LArTPC. 
A charged track in the TPC volume ionizes electrons.  
An imposed electric field (500 V/cm) causes the ionization to 
drift toward readout anode wire planes and be collected. 
The signals from the wire planes provide 2 dimensional information about the event. 
The drift time of the ionization electrons gives the third coordinate. 
In other words, the time information is ``projected'' onto the third dimension. 
The combined wire and time information allows the tracks to be reconstructed in three dimensions. 
(Right) A drawing of the MicroBooNE cryostat with the TPC and PMT arrays on the left wall. 
}
\label{fig:LArTPC}
\end{figure}

Since the LArTPC was first proposed~\cite{Rubbia}, the detection technique 
has mainly been
developed by an Italian collaboration~\cite{ICARUS_T600}. Its physics potential, 
especially in neutrino interaction measurements, 
has been demonstrated~\cite{ICARUS_MILANO}. 
The notable features of this detector are its three-dimensional and calorimetric 
reconstruction of charged particle tracks 
(Figure~\ref{fig:LArTPC}, left). 

The first US LArTPC to take data in a neutrino beamline is the ArgoNeuT detector~\cite{Josh}. 
It has the advantage of the MINOS near detector~\cite{MINOS_ND}, 
located behind of the ArgoNeuT detector, as a muon range stack. 
Table~\ref{tab:LArTPC} summarizes the ArgoNeuT detector parameters. 
The ArgoNeuT experiment is discussed further in these proceedings~\cite{Josh}.

\subsection{MicroBooNE --- Motivation}

Although ArgoNeuT mainly focuses on detector R\&D,  
MicroBooNE features 50\% R\&D and 50\% physics (Figure~\ref{fig:LArTPC}, right). 
In a LArTPC, single electron tracks can be distinguished from gamma rays that have 
converted to e$^+-$e$^-$ pairs by 
utilizing the conversion length of the gamma ray and $\frac{dE}{dx}$ of the electron. 
Thus, MicroBooNE should have excellent particle identification for 
$\nu_e$ ($\bar\nu_e$) appearance oscillation experiments, 
where gamma rays are backgrounds. 
This discrimination power can be used to study the MiniBooNE low energy excess~\cite{MB_unexp},
which is not currently understood. 
If this excess is made by misidentification of a gamma ray, either from a known or unknown source, MicroBooNE can reject them. 

\subsection{Detector R\&D}

\subsubsection{Fermilab materials test system (MTS)}

\begin{figure}
\includegraphics[height=2.3in]{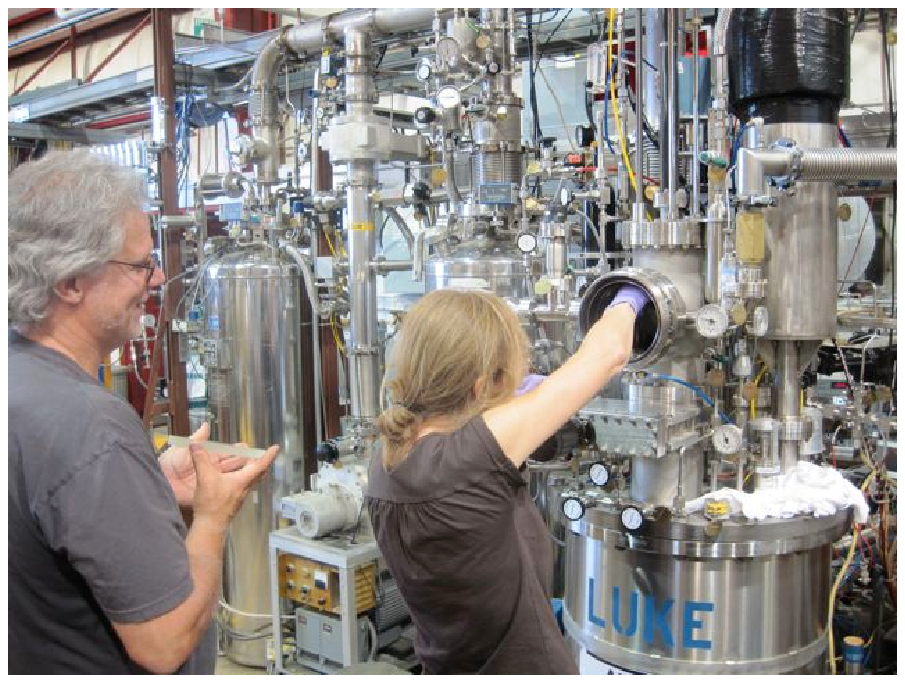}
\includegraphics[height=2.3in]{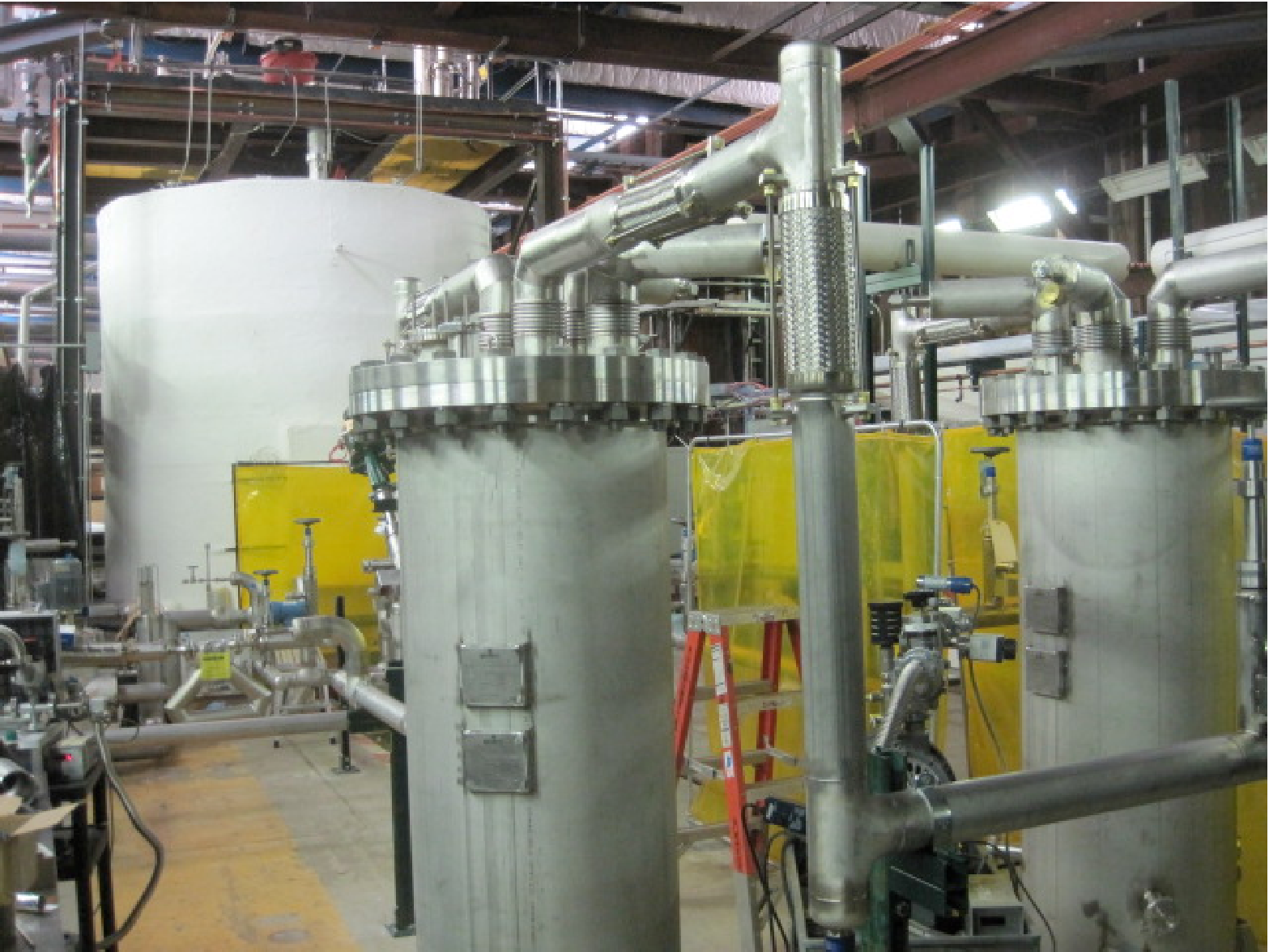}
\caption{
(color online). (Left) The Fermilab MTS cryostat, ``LUKE''. 
In this picture, a scientist prepares a test sample in the sample cage. 
(Right) The Liquid Argon Purity Demonstrator (LAPD). 
2 Zeolite filters are seen in front and the 30 ton vessel 
is seen in back.
}
\label{fig:LUKE}
\end{figure}

Table~\ref{tab:LArTPC} summarizes the MicroBooNE parameters.  
This bigger detector guarantees more neutrino interactions, 
although it also requires a longer drift length of ionized electrons 
in order to see the charged particle tracks. As such, 
the liquid Argon is required to have a higher purity. 
For MicroBooNE, to preserve $>$50\% of electrons to be drifted 
without attaching to electro-negative impurities, 
we need to achieve $\sim 100$~ppt level of 
oxygen-equivalent impurity concentration~\cite{filter}. 
For this goal, we have several test facilities. 
The Fermilab MTS cryostat ``LUKE'' (Figure~\ref{fig:LUKE}, left) 
has a window at the airlock region. 
The test material is inserted and sample cage is lowered into the LUKE volume, 
which is filled with high purity liquid Argon. 
The impurity is monitored in the Argon gas and liquid regions. 
All materials used inside of the MicroBooNE cryostat are 
required to be tested by MTS.

\begin{table} 
\begin{tabular}{lccc} 
\hline
                 & ArgoNeuT   & MicroBooNE & LAr20       \\ 
\hline 
Cryostat volume  & 0.7~ton    & 150~ton    & 25,000~ton  \\
TPC volume       &0.25~ton    &  86~ton    & 16,800~ton  \\
Max. drift length& 0.5~m      & 2.5~m      &    2.5~m    \\
Electronics      &JFET (293~K)&CMOS (87~K) &CMOS (87~K)  \\
\# of channel    &   480      &$\sim$9,000 &$\sim$645,000\\
Wire pitch       &   4~mm     &    3~mm    &    3~mm     \\
\# of wire plane &   2        &    3       &     3       \\
Light collection & none       &30 of 8''PMT&    TBD      \\
\hline 
\end{tabular} 
\caption{Summary of 3 US LArTPCs. The R\&D started from 
the modestly sized ArgoNeuT experiment. The next goal is to conduct the MicroBooNE experiment. 
The ultimate goal is a large kiloton-scale LArTPC, such as LAr20.} 
\label{tab:LArTPC} 
\end{table} 

\subsubsection{Liquid Argon Purity Demonstrator (LAPD)}

MicroBooNE is one of the key steps on the path to a future large 
LArTPC detector, such as LAr20 (Tab.~\ref{tab:LArTPC}), 
a candidate detector of long baseline neutrino experiment (LBNE)~\cite{LBNE}. 
Presumably, we cannot evacuate an ultra-large cryostat  
before filling it with liquid Argon in order to remove impurities. 
LAPD (Figure~\ref{fig:LUKE}, right) 
is a 30 ton vessel with 2 copper (Oxygen) and 2 Zeolite (water) filters. 
We are testing ways to ``push out'' all impurities inside the vessel 
by flowing Argon gas through it, rather than evacuating the whole vessel. 
The purity and temperature are monitored at several places inside the tank. These 
provide important inputs for the design of a large LArTPC.


\subsubsection{Cryogenic large PMT system}

\begin{figure}
\includegraphics[height=2.3in]{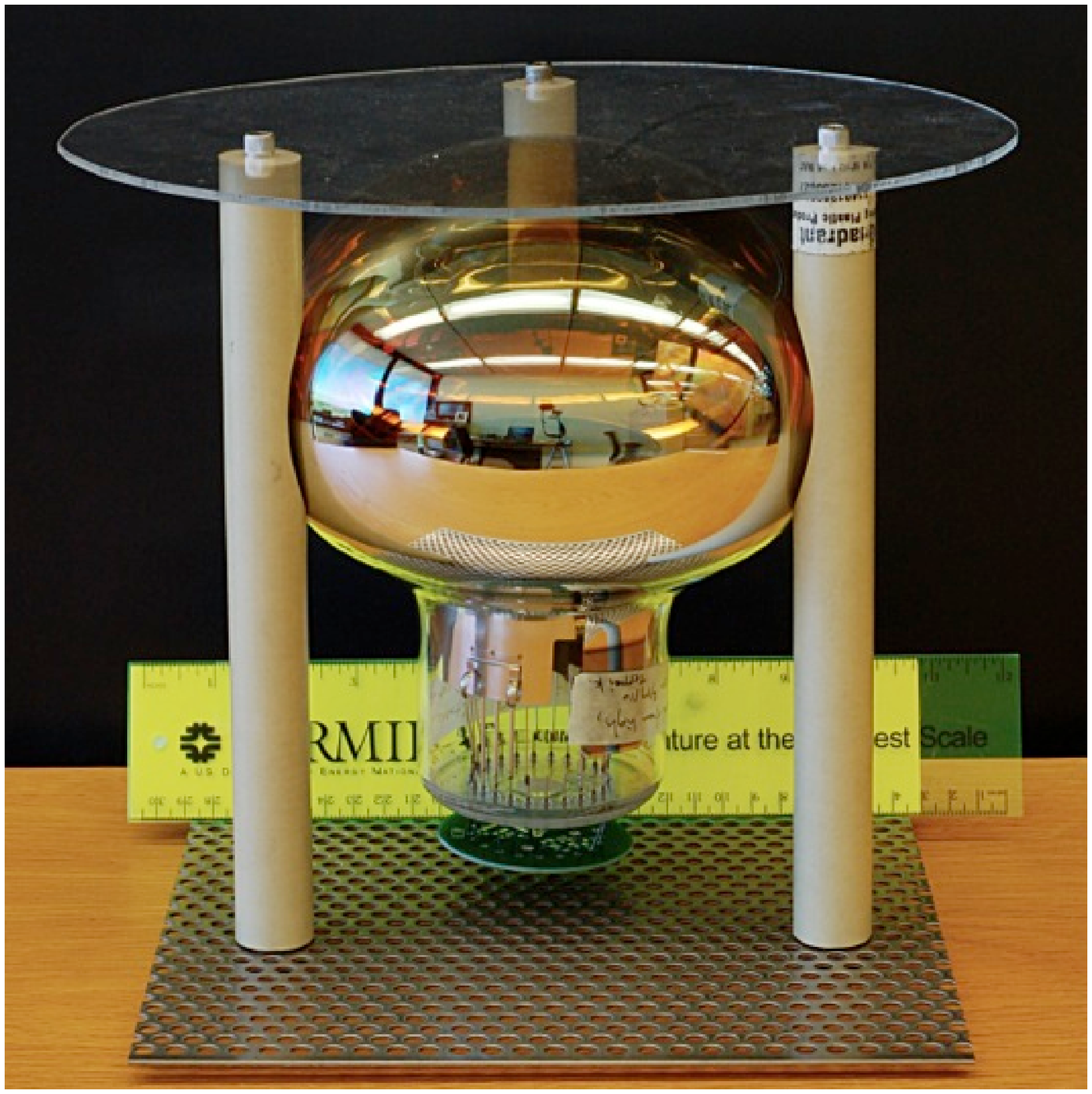}
\includegraphics[height=2.3in]{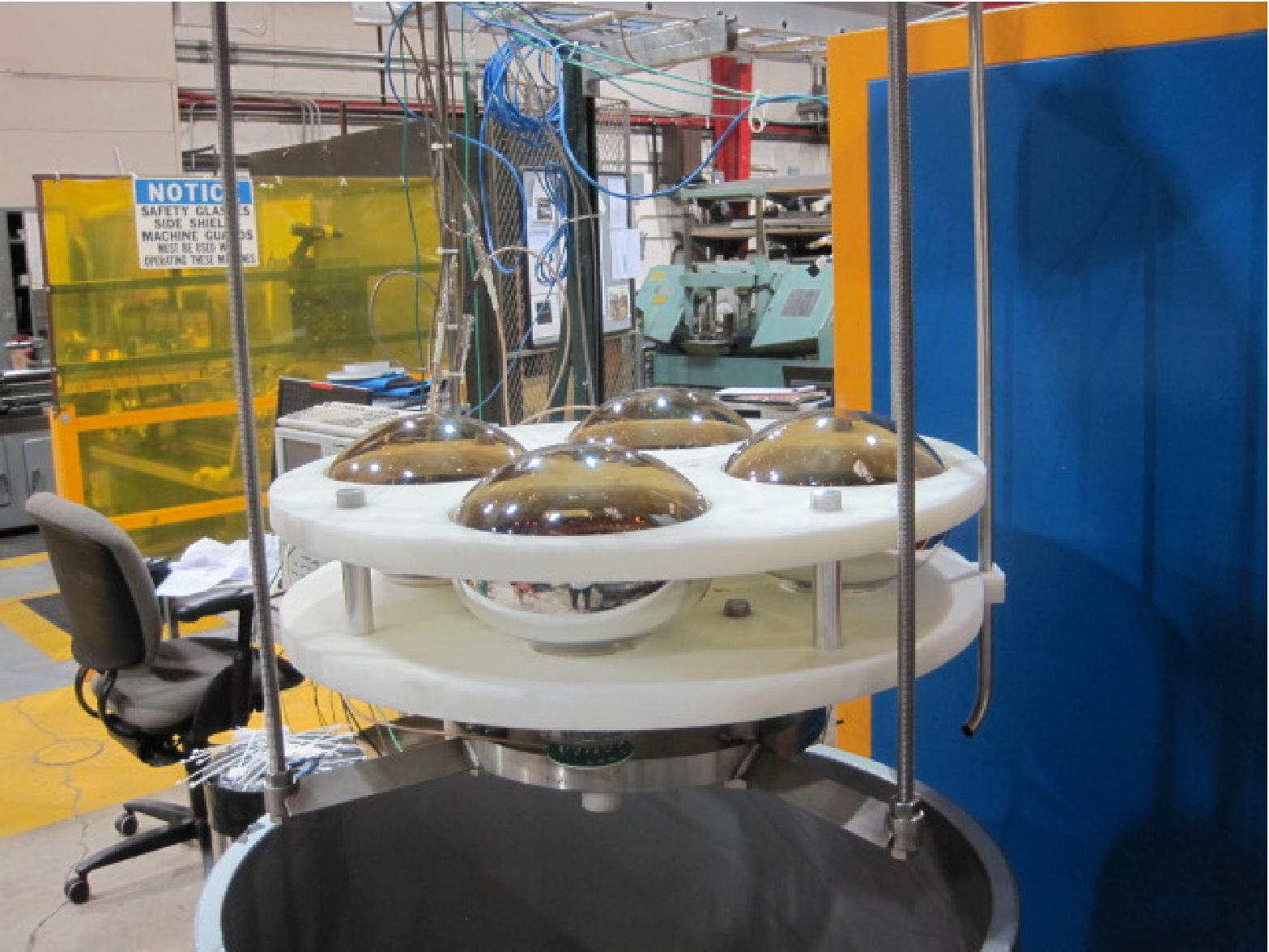}
\caption{
(color online). (Left) A mechanical model of a PMT unit of MicroBooNE. 
A 8'' cryogenic PMT is supported by 3 PEEK posts with a wave length shifting plate on top. 
The wave length shifting plate is an acrylic plate, 
coated with a 50\%-50\% mixture of Tetra-Phenyl butadiene (TPB) 
and polystyrene as a wave length shifter. 
TPB shifts 128nm VUV scintillation light into the blue region, 
where the bi-alkali photo-cathode is the most sensitive. 
(Right) The PMT test stand at Fermilab Proton Assembly Building (PAB). 
4 PMTs are simultaneously immersed in liquid nitrogen for testing. 
}
\label{fig:PMT}
\end{figure}

MicroBooNE will employ 30 8'' cryogenic PMTs. 
Figure~\ref{fig:PMT} (left) shows the PMT unit mechanical model.  
Since the prompt component of scintillation light is much faster (6~ns) than 
the electron drift velocity of the TPC (1.6~m/ms), scintillation light 
can be used as a trigger for the TPC detector.
These PMTs have Platinum coating under the photo-cathode 
that works below 150 K. 
The PMT base is custom designed in order to function in the cold environment. 
The vacuum UV (VUV) scintillation light from the liquid Argon 
is shifted to blue region by the wave length shifting plate.
Figure~\ref{fig:PMT} (right) is a picture of our set up of the 
PMT test stand at Fermilab PAB. 

\subsection{Physics}

Table~\ref{tab:event} shows the summary of the predicted 
event rate for MicroBooNE by 
NUANCE neutrino interaction generator~\cite{NUANCE}. 
On the order of 10,000 interactions 
are expected in many channels. 

\begin{table}  
\begin{tabular}{llc} 
\hline 
production mode & formula &\#evt ($\times 10^3$)\\ 
\hline 
CC quasi-elastic        &$\numu+n\to\mu^-+p$          & 66\\
NC elastic              &$\numu+N\to\numu+N$          & 21\\
CC resonance $\pi^+$    &$\numu+N\to\mu^-+N+\pi^+$    & 28\\
CC resonance $\pi^\circ$&$\numu+n\to\mu^-+p+\pi^\circ$&  7\\
NC resonance $\pi^\circ$&$\numu+N\to\numu+N+\pi^\circ$&  8\\
NC resonance $\pi^\pm$  &$\numu+N\to\numu+N'+\pi^\pm$ &  3\\
CC DIS                  &$\numu+N\to\mu^-+X, W>$2~GeV &  1\\
NC DIS                  &$\numu+N\to\numu+X, W>$2~GeV &0.5\\
CC coherent $\pi^\circ$ &$\numu+A\to\mu^-+A+\pi^+$    &  3\\
NC coherent $\pi^\pm$   &$\numu+A\to\numu+A+\pi^\circ$&  2\\
CC Kaon production      &$\numu+N\to\mu^-+K+X$        &$\sim$0.1\\
NC Kaon production      &$\numu+N\to\numu+K+X$        &$<$0.1\\
others                                               &&  4\\
\hline
total                                                &&143\\
\hline 
\end{tabular} 
\caption{Summary of event rates in MicroBooNE's 60 ton volume by 
the neutrino mode BNB (6E20 POT), 468~m from the target.} 
\label{tab:event} 
\end{table} 

\subsubsection{Neutral Current Elastic (NCE) measurement and $\De$s}

Neutral current elastic (NCE) scattering is uniquely sensitive 
to the iso-scalar part of the axial current form factor, 
which is generally related with $\De$s, 
the strange quark spin component of the nucleon. 
$\De$s relates the elastic scattering 
form factor and the parton distribution function (PDF), 
\beq
G_A^s(Q^2=0)=\De s=\int_0^1 \De s(x)dx~,\no
\eeq
which is measured through semi-inclusive DIS experiments~\cite{HERMES}. 
However, the obtained values disagree with $\De$s from neutrino 
NCE measurement~\cite{BNLE734}. 
To determine $\De$s, measurements of low energy protons are crucial.  
MicroBooNE will be able to resolve very short proton tracks 
(a design goal is to measure $\sim$1.5~cm, 
equivalent to $\sim$40~MeV kinetic energy). 
Such tracks are unmeasurable with Cherenkov~\cite{MB_NCE} 
and fine-grained tracking detectors~\cite{SB_NCE}. 


\subsubsection{Upgraded NuMI (Neutrinos at the Main Injector) beam}

MicroBooNE will be a BNB on-axis experiment. 
However, it will receive on the order of 40,000 $\numu$CC 
and 7,700 $\numubar$CC interactions per year from the 
upgraded NuMI beam for the NOvA experiment~\cite{NOvA}, 
where the beam is expected to run with 700kW ($\sim$6E20 POT/year) power. 
NuMI will provide a higher flux along with a surplus of anti-neutrinos. 
NuMI is the only source of $\numubar$ since the 
BNB is currently not planned to run with anti-neutrino mode (and BNB is 93\% pure $\numu$).  
The high $\numubar$ flux is an ideal place to study hyperon production, 
\beq
\numubar+p\to\mu^++\La^\circ~,~\numubar+p\to\mu^++\Si^\circ~.\no
\eeq 
We expect 100s of events per year (before cuts), 
which is much more than the integrated world data~\cite{Lambda}. 
The angular distribution measurement of decay products allow
the study of $\La^\circ$ polarization in neutrino interactions. 

\subsubsection{Neutrino two nucleon short range correlation ($\nu$2NSRC) measurement}

The nucleon correlation is a recent hot topic in nuclear physics.  
A series of recent electron scattering experiments, mainly from JLab, 
show an emerging view, namely, 
(1) for a heavy nuclei, $\sim$20\% of nucleons are in a 2NSRC state~\cite{CLAS}, 
(2) 90\% are proton-neutron pair~\cite{Piasetzky}, and 
(3) 2 ejected nucleons are kinematically correlated~\cite{HALLA}. 
In a $\numu$CC interaction, this interaction signature is a triple coincidence; 
a muon with 2 correlated protons, 
\beq
\numu+X(n-p)\to\mu^-+p+p+X'~. \no
\eeq 
The $\nu$2NSRC is interesting for neutrino scattering in three different ways. 
First, 2NSRC is not confirmed to occur in neutrino scattering~\footnote{
There is an indication of $\nu$2NSRC in old bubble chamber experiment, see ~\cite{Roe}}.  
Second, nuclear break-up by 2NSRC is not studied via electron scattering, 
because typical electron scattering experiments use long arm spectrometers~\cite{EliMark}. 
On the other hand, neutrino experiments use vertex detectors, 
so neutrino experiments can measure energy release from the vertex. 
Third, there is wide-spread speculation that $\nu$2NSRC may contribute to recent 
large neutrino cross section results from MiniBooNE~\cite{MB_NCE,MB_others}, 
SciBooNE~\cite{SB_others}, and MINOS~\cite{MINOS}. 
All modern neutrino experiments have measured 
higher cross sections than have been predicted, 
except NOMAD~\cite{NOMAD} and MINERvA preliminary result~\cite{MINERvA}. 
In the theories~\cite{Martini,Nieves,Barbaro}, 
2NSRC enhances generally hard to identify multi-nucleon emission channels. 
However, high resolution LArTPC has the ability to track protons, and 
it has the potential to identify $\nu$2NSRC for the first time in history.  
This process may shed light on neutrino cross section measurements around 1~GeV. 

%

\bibliographystyle{aipproc}

\end{document}